\documentclass[twocolumn]{aastex63}

\usepackage{amsmath}
\usepackage{graphicx}
\usepackage{epstopdf}
\usepackage{enumerate}
\usepackage{soul}

\usepackage{savesym}
\savesymbol{tablenum}
\usepackage{siunitx}
\restoresymbol{SIX}{tablenum}

\usepackage{hyperref}
\usepackage{orcidlink} 

\newcommand{\beq}{\begin{equation}}
\newcommand{\eeq}{\end{equation}}
\newcommand{\beqn}{\begin{eqnarray}}
\newcommand{\eeqn}{\end{eqnarray}}
\newcommand{\pa}{\partial}

\begin{document}


\title{Threshold mass of the general relativistic instability for supermassive star cores}

\author[0000-0002-4979-5671]{Masaru Shibata}
\affiliation{Max-Planck-Institut f\"ur Gravitationsphysik (Albert-Einstein-Institut), Am M\"uhlenberg 1, D-14476 Potsdam-Golm, Germany}
\affiliation{Center for Gravitational Physics and Quantum Information, Yukawa Institute for Theoretical Physics, Kyoto University, Kyoto, 606-8502, Japan}

\author[0000-0001-6467-4969]{Sho Fujibayashi}
\affiliation{Frontier Research Institute for Interdisciplinary Sciences, Tohoku University, Sendai 980-8578, Japan}
\affiliation{Astronomical Institute, Graduate School of Science, Tohoku University, Sendai 980-8578, Japan}
\affiliation{Max-Planck-Institut f\"ur Gravitationsphysik (Albert-Einstein-Institut), Am M\"uhlenberg 1, D-14476 Potsdam-Golm, Germany}

\author[0009-0007-7617-7178]{C\'edric Jockel}
\affiliation{Max-Planck-Institut f\"ur Gravitationsphysik (Albert-Einstein-Institut), Am M\"uhlenberg 1, D-14476 Potsdam-Golm, Germany}

\author[0000-0003-4443-6984]{Kyohei Kawaguchi}
\affiliation{Max-Planck-Institut f\"ur Gravitationsphysik (Albert-Einstein-Institut), Am M\"uhlenberg 1, D-14476 Potsdam-Golm, Germany}
\affiliation{Center for Gravitational Physics and Quantum Information, Yukawa Institute for Theoretical Physics, Kyoto University, Kyoto, 606-8502, Japan}
\affiliation{Institute for Cosmic Ray Research, The University of Tokyo, 5-1-5 Kashiwanoha, Kashiwa, Chiba 277-8582, Japan}





\date{\today}


\begin{abstract}
The dependence of the final fate of supermassive star (SMS) cores on their mass and angular momentum is studied with simple modeling. SMS cores in the hydrogen burning phase encounter the general relativistic instability during the stellar evolution if the mass is larger than $\sim 3 \times 10^4M_\odot$. Spherical SMS cores in the helium burning phase encounter the general relativistic instability prior to the onset of the electron-positron pair instability if the mass is larger than $\sim 1\times 10^4M_\odot$. For rapidly rotating SMS cores, these values for the threshold mass are enhanced by up to a factor of $\sim 5$, and thus, for SMSs with mass smaller than $\sim 10^4M_\odot$ the collapse is triggered by the pair-instability, irrespective of the rotation. After the onset of the general relativistic instability, SMS cores in the hydrogen burning phase with reasonable metallicity are likely to collapse to a black hole irrespective of the degree of rotation, whereas the SMS cores in the helium burning phase could explode via nuclear burning with no black hole formation, as previous works demonstrate. 
\end{abstract} 



\section{Introduction}\label{secI}

A supermassive star (SMS) is a possible progenitor for the formation of a seed of a supermassive black hole (SMBH). Recent star-formation calculations with high-mass accretion rates in spherical symmetry \citep{2013ApJ...778..178H,2016ApJ...830L..34U,2018MNRAS.474.2757H} suggest that if a high mass-accretion rate of $\agt 0.1M_\odot$/yrs is preserved during the nuclear burning phases of $\sim 2 \times 10^6$\,yrs, a SMS with mass $\agt 10^5M_\odot$ could be formed. Subsequently, the SMS core may collapse to a seed of a SMBH with mass $\agt 10^4M_\odot$. 
Because of their high mass, the general-relativistic radial instability~\citep{1963ApJ...138.1090I,1964ApJ...140..417C,1971reas.book.....Z,1983bhwd.book.....S,1986ApJ...307..675F} is often referred to as the mechanism of the collapse of the SMS cores. On the other hand, for massive stars with mass $\alt 10^4M_\odot$, the pair instability is likely to be the mechanism for the onset of the gravitational collapse (see, e.g., \cite{1971reas.book.....Z, 1983bhwd.book.....S, 1996snih.book.....A}). However, to our knowledge, the threshold mass, in particular for rotating SMS cores, is not yet well understood. In the following, we pay particular attention to the fate of rotating SMS cores with a variety of masses and chemical compositions: We explore the SMS cores not only in the hydrogen burning phase but also in the helium burning phase. 

The mechanism that induces the collapse to a massive black hole plays an important role for determining the subsequent outcome. In our case, we consider the general relativistic instability. There, the unstable mode is the fundamental radial mode with no nodes. In other words, the displacement vector of the Lagrangian perturbation of the unstable mode is approximately proportional to $r$, where $r$ is the radial coordinate~\citep{1964ApJ...140..417C}. Thus, the collapse associated with this instability proceeds in a coherent manner, i.e., all the stellar matter starts collapsing simultaneously. On the other hand, the pair instability occurs when the adiabatic index decreases below $4/3$ by the depletion of the thermal pressure due to the pair creation of electron-positron ($e^-e^+$) pairs~\citep{1996snih.book.....A}. This initially happens only in the central region of stars. The collapse therefore does not proceed as coherently as for the general relativistic instability; the central region collapses first, and then the outer region follows due to pressure depletion.

After the collapse of a rotating SMS core, a rotating black hole is expected to be formed~\citep{2002ApJ...577..904S,2003ApJ...595..992S,2016ApJ...830L..34U}. A disk/torus can then be formed around the newly formed black hole~\citep{2002ApJ...572L..39S,2007PhRvD..76h4017L,2012ApJ...749...37M,Uchida2017oct,2017PhRvD..96d3006S}. This is because the infalling material from the outer part of the SMS core has sufficient specific angular momentum. Due to the coherent nature of the collapse after encountering the general relativistic instability, a strong shock is generated at the surface of the disk/torus formed around the black hole~\citep{2007PhRvD..76h4017L,Uchida2017oct} (see also \citealt{2014ApJ...792...44C,2020MNRAS.496.1224N,Nagele2022dec,Nagele:2024aev} for other explosion channels with a specific mass range of helium-burning SMSs). This shock subsequently induces a mass ejection similar to a supernova explosion with high kinetic energy of $\sim 10^{-5}$--$10^{-4}M c^2$, where $M$ is the mass of the SMS core and $c$ is the speed of light. By contrast, such a strong explosion from the disk formation is not seen in the case of a pair-instability collapse (see, e.g., \citealt{2019PhRvD..99d1302U} for a fully general relativistic simulation). 


In recent years, the James-Webb Space Telescope (JWST) has observed a number of high luminosity SMBH candidates in the early universe~(e.g., \citealt{2023ApJ...957L...3P,2023ApJ...955L..24G,2023ApJ...959...39H, 2023ApJ...954L...4K, 2024arXiv240405793M,Kovacs:2024zfh,Bogdan:2023ilu}). It is usually believed that such highly luminous objects shine as a result of rapid mass accretion onto a SMBH. One alternative possibility, although it is unlikely to be the majority, is the explosion soon after the onset of the collapse of a SMS core. As we mentioned above, the explosion energy for a SMS core with $M=10^5M_\odot$ can be $\sim 10^{55}$--$10^{56}$\,erg. The explosion is accompanied by significant mass ejections. Since the diffusion timescale of photons is quite long, the duration of the supernova-like phenomena can be $\sim 10$\,yrs~\citep{Uchida2017oct,Matsumoto:2015bjg,Matsumoto_2016}. This implies that at a high cosmological redshift of $z \sim 10$, the observed duration can be up to $\sim 100$\,yrs. \cite{Uchida2017oct} also inferred that the peak luminosity for such an explosion can be $10^{45}$\,erg/s for the collapse of rapidly rotating SMS cores, which is comparable to the recently observed luminous SMBH candidates.

A word of caution may be appropriate here. Since SMSs shine with nearly Eddington luminosity, mass accretion with high angular velocity may be prevented by the radiation pressure. This limit on the rotation rate due to combined effects of centrifugal force and radiation pressure is called the $\Omega\Gamma$-limit, see \cite{Lee_2016,2018ApJ...853L...3H}. SMSs may therefore only rotate slowly. However, the previous stellar evolution calculations were based essentially on the assumption of spherical symmetry. Hence, non-spherical effects on the stellar profile and radiation field are not accurately taken into account. Indeed a latest star formation simulation~\citep{2023ApJ...950..184K} shows that the radiation field is non-spherical for rapidly rotating systems and the radiation pressure on the equatorial direction is appreciably weaker than that along the polar axis, allowing for the spin-up of the massive star. Moreover the $\Omega\Gamma$-limit might not be important for the core region of SMSs, if SMSs were differentially rotating with a rapidly rotating core, close to the Kerpler limit. Indeed this possibility is suggested by sellar evolution simulations of SMSs \citep{2018ApJ...853L...3H}. We therefore focus on studying SMS cores and explore not only slowly rotating but also rapidly rotating ones. For simplicity we also assume that SMS cores are rigidly rotating. This assumption is likely to be valid at least for the convective region, of which SMS cores are largely composed.

The paper is organized as follows. In Sec.~\ref{sec2}, we review the equation of state for SMSs. We pay particular attention to the dependence of their mass on the equation of state and the stellar composition. Section~\ref{sec3} presents a variety of SMS core models in the hydrogen and helium burning phases. The computation is performed in the framework of general relativity because we are interested in the general relativistic instability. We determine the threshold mass for the onset of the general relativistic instability and clarify its dependence on the rotational angular velocity. We also show that stable sequences of the SMS cores in the helium burning phase are unlikely to encounter the pair instability unless the mass fraction of helium is extremely low ($\alt 10^{-6}$). In Sec.~\ref{sec4}, we briefly analyze for which case the explosion of SMSs can be driven by explosive nuclear burning after the onset of the general relativistic instability. Section~\ref{sec5} is devoted to a summary. Throughout this paper, $c$, $G$, $a_r$, and $k_B$ denote the speed of light, gravitational constant, radiation constant, and Boltzmann's constant, respectively. 

\section{Equations of state}\label{sec2}

In this paper we suppose that SMS cores are composed of hydrogen, helium, carbon, oxygen, electrons, and photons. We also assume that they are in a nuclear burning phase of hydrogen or helium. Here, the helium burning implies the triple alpha reaction, the reaction between helium and carbon, and between helium and oxygen (but we do not consider the presence of the remnant of the helium-oxygen reaction for simplicity). Then, the pressure, $P$, and internal energy density, $\epsilon$, are written as~\citep{Bond:1984sn}
\beqn
P &=&{a_r T^4 \over 3} + Y_T n k_B T,\label{pressure}\\
\epsilon&=&a_r T^4 + {3 \over 2} Y_T n k_B T,\label{epsilon}
\eeqn
where $T$ is the temperature and $n$ is the baryon number density. $Y_T$ is the total number of particles per baryon, defined by
\beq
Y_T \equiv Y_e + Y_\mathrm{p} + Y_\mathrm{He} + Y_\mathrm{C} + Y_\mathrm{O},
\eeq
where $Y_I=n_I/n$ with $n_I$ for $I =\{e, p,\mathrm{He}, \mathrm{C}, \mathrm{O}\}$ denoting the number density of electron, hydrogen, helium, carbon, and oxygen, respectively. For simplicity we approximately set the mass fraction of hydrogen, helium, carbon, and oxygen as $X_\mathrm{p}=Y_\mathrm{p}$, $X_\mathrm{He}=4Y_\mathrm{He}$, $X_\mathrm{C}=12Y_\mathrm{C}$, and $X_\mathrm{O}=16Y_\mathrm{O}$. For primordial gas, $X_\mathrm{p} \approx 0.75$, $X_\mathrm{He} \approx 0.25$, and $Y_e \approx 0.875$, and thus, $Y_T \approx 1.69$. During the stellar evolution by nuclear burning, the value of $Y_T$ monotonically decreases.

For SMSs in nuclear burning phases, in particular for their core region, convection should be highly enhanced, and a convective equilibrium should be realized \citep{Bond:1984sn,2016ApJ...830L..34U}. This implies that the SMS cores are approximately isentropic, i.e., the entropy per baryon, $s$, is constant. The chemical composition of the SMS core is therefore approximately uniform, i.e., $Y_T=$const. 
Although this assumption is a significant simplification, it enables us to determine equilibrium states for rotating SMS cores while taking into account their structure accurately. We note that this assumption is obviously incorrect in the late stage of stellar evolution, in which the heavier elements are located in the central region and thus $s$ and $Y_T$ are not uniform (e.g., \citealt{2020MNRAS.496.1224N}). Moreover a shell burning can be the major nuclear reaction~\citep{1990sse..book.....K} in such a stage. Computation of non-spherical stellar structure for non-uniform values of $s$ and $Y_T$ is one of the challenging issues in future. 

Under the condition of $s=$const, the first law of thermodynamics, $d(\epsilon/n)=-Pd(1/n)$, gives the relation between $dT$ and $dn$ (i.e., between $dP$ and $dn$) from Eqs.~(\ref{pressure}) and~(\ref{epsilon}). Then the adiabatic constant is written as \citep{1939isss.book.....C,Bond:1984sn}
\beq
\Gamma=\left({\pa \ln P \over \pa \ln n}\right)_{s}
={4 \over 3}+{4\sigma + 1 \over 3(\sigma+1)(8\sigma+1)},
\eeq
where $\sigma$ is the ratio of the radiation pressure to the gas
pressure defined by
\beq
\sigma \equiv {a_r T^3 \over 3Y_T n k_B}={s_\gamma \over 4Y_T k_B},
\label{sigmadef}
\eeq
and $s_\gamma$ here denotes the photon entropy per baryon.  For SMS cores with $\Gamma-4/3 \gtrsim 0$, $\sigma$ and $s_\gamma/k_B$ are much larger than unity, and thus, $\Gamma$ can be approximated by $\Gamma \approx 4/3 + 1/(6\sigma)$. Throughout this paper, we will use this approximate relation. 

Because the photon entropy is much larger than the gas entropy and $s$ is assumed to be constant, we may also assume that $s_\gamma$ and $\sigma$ are approximately constant.  Hence, it is reasonable to assume that the equation of state for the SMS cores are well approximated by a polytropic form
\beq
P=K\rho^\Gamma,~~~~\Gamma=1+{1 \over n_p}, \label{poly}
\eeq
where $\rho$ is the rest-mass density ($\rho=m_B n$ with $m_B$ the mean baryon mass). $K$ and $n_p$ are the adiabatic constant and polytropic index, respectively. 

Using Eqs.~(\ref{pressure}), (\ref{sigmadef}), and (\ref{poly}), the polytropic constant may be written as
\beqn
K=\left({Y_T k_B \sigma \over m_B}\right)^{4/3}
\left({3 \over a_r}\right)^{1/3}
\left(1+\sigma^{-1}\right)\rho^{-1/(6\sigma)}. 
\eeqn
For $\sigma \gg 1$, $K$ is considered to be constant throughout the SMS core.
From $K$, we can construct the quantity of mass dimension as
\beq
M_u \equiv K^{n_p/2} G^{-3/2}c^{3-n_p}.\label{massu}
\eeq
For $\sigma \gg 1$, this quantity is written as
\beq
M_u=M_{u,3}\left(1+{3 \over 2\sigma} \right)
\left({m_B c^2 \over Y_T k_B T \sigma}\right)^{3/(4\sigma)},
\label{massunit}
\eeq
where $M_{u,3}$ denotes $M_u$ for $n_p=3$ ($\Gamma=4/3$) and is written as
\beqn
M_{u,3}&=&\left({Y_T k_B \over m_B}\right)^2
\left({3 \over G^3 a_r}\right)^{1/2} \sigma^2  \nonumber \\
&\approx& 4.01M_\odot Y_T^2 \sigma^2 . \label{mass20} 
\eeqn
To derive Eq.~(\ref{massunit}), we used Eq.~(\ref{sigmadef}). Equation~(\ref{mass20}) shows that the mass of a SMS core is approximately proportional to $(Y_T \sigma)^2$. During the stellar evolution by nuclear burning, $Y_T$ decreases monotonically with the increase of the average mass number of the elements in the SMS core. Assuming that the mass of the SMS core is approximately constant during the evolution, $\sigma$ increases, i.e., $\Gamma$ decreases, with the stellar evolution. Therefore, during the evolution of SMS cores, they can become more susceptible to the general relativistic instability (cf. Sec.~\ref{secIIID}). 

\section{Equilibrium states of supermassive star cores and stability} \label{sec3}

\subsection{Basic Equations}

Assuming that SMS cores are composed of an ideal fluid, we write the stress-energy tensor as
\beqn
T^{\mu\nu}=\rho h u^{\mu} u^{\nu} + P g^{\mu\nu},
\eeqn
where $u^{\mu}$ is the four velocity, $h \equiv c^2 + \varepsilon + P/\rho$, the specific enthalpy, $\varepsilon$ the specific internal energy (different from $\epsilon$), and $g_{\mu\nu}$ the spacetime metric.  As we already mentioned in Sec.~\ref{sec2}, we use the polytropic equation of state. Together with the first law of thermodynamics, $\varepsilon$ can then be written as
\beq
\varepsilon={n_p P \over \rho}. 
\eeq
As also mentioned in Sec.~\ref{sec2}, we assume that SMS cores are in convective equilibrium. Previous works~(e.g., \citealt{1999ApJ...526..941B}) have suggested that a turbulent state is realized by the convection inside SMS cores, generating an effective viscosity. Angular velocity and the entropy per baryon would then be approximately uniform. Thus, we pay attention only to rigidly rotating SMS cores and set the angular velocity $\Omega \equiv u^{\varphi}/u^t$ to be constant. Under this set-up, the Euler equation is integrated to give the first integral, and hence, the procedure for the equilibrium solutions becomes straightforward~\citep{1976ApJ...204..200B,1986ApJ...304..115F}. By contrast, if the isentropic condition is not satisfied or if a meridional circulation is assumed to be present, we have to solve the Euler equation directly. 


For the spherical symmetric case with $\Omega=0$, we solve the Tolman-Oppenheimer-Volkoff equations~\citep{1939PhRv...55..364T,1939PhRv...55..374O}. For the rotating case, we follow \cite{1976ApJ...204..200B} and write the line element as
\beqn
ds^2 &=& -e^{2\nu} c^2 dt^2
+ B^2 e^{-2\nu} r^2 \sin^2\theta (d\varphi - \omega c dt)^2 \nonumber \\
&& +e^{2\zeta-2\nu}(dr^2 + r^2 d\theta^2), 
\eeqn
where $\nu$, $B$, $\omega$, and $\zeta$ are field functions of $r$ and $\theta$. The first three obey elliptic-type equations in axial symmetry, and the last one an ordinary differential equation. These equations are solved using the method of \cite{1998PhRvD..58j4011S}.  


The rest mass $M_*$, Komar mass (gravitational mass) $M$, Komar angular momentum $J$, rotational kinetic energy $T_{\rm rot}$, gravitational potential energy $W$, and internal energy $U$ are defined by
\beqn
M_* &=& 2\pi c \int \rho u^t B e^{2\zeta-2\nu} r^2 dr d(\cos\theta), \\ 
M &=&2\pi c^{-2}\int (-2T_t^{~t}+T_{\mu}^{~\mu}) B e^{2\zeta-2\nu}
r^2 dr d(\cos\theta),~~~~%
\\
J  &=&2\pi c^{-1} \int \rho h u^t u_{\varphi} B e^{2\zeta-2\nu}
r^2 dr d(\cos\theta),\\ 
T_{\rm rot}&=&{1 \over 2} J \Omega,
\\
W&=& M_{\rm p} - M + T_{\rm rot} ~(>0), \\
U&=&M_\mathrm{p}-M_*,
\eeqn
where $M_\mathrm{p}$ is the proper mass defined by
\beq
M_{\rm p} =2\pi \int \rho u^t (1+c^{-2}\varepsilon)
B e^{2\zeta-2\nu} r^2 dr d(\cos\theta). 
\eeq
From these quantities, the dimensionless parameters associated with rotation are defined as 
\beqn
\beta:={T_{\rm rot} \over W}~~\mathrm{and}~~
q:={cJ \over GM^2}.  \label{eq:beta-rotation-parameter}
\eeqn


In addition to these quantities, we often refer to the central density, $\rho_\mathrm{c}$, which is used to specify a rotating star for a given set of $\beta$ and $M$. We also consider the equatorial circumferential radius, $R_\mathrm{e}$, by which a compactness parameter is defined by 
\beq
{\cal C}:={GM \over c^2R_\mathrm{e}}. 
\eeq
Using the central pressure $P_\mathrm{c}$, we also refer to a dimensionless quantity, 
\beq
y:={P_\mathrm{c} \over \rho_\mathrm{c} c^2}, \label{eq:dimensionless-quantity-y}
\eeq
for specifying the compactness of rotating SMS cores, which has the same order of magnitude of $GM/(c^2R_e)$. For $n_p=3$ spherical polytropes in Newtonian gravity, the value of $P_c/\rho_c$ is equal to $(GM/R)/1.1705$ where $R$ is the stellar radius. 

We note that real SMSs are likely to have high matter accretion rates~\citep{2013ApJ...778..178H}. Due to this accretion, a bloated convective envelope will form. This envelope will have a stellar radius (photospheric radius) that is $\agt 10$ times larger than the core region, but have a small total mass compared to the SMS core (see, e.g., \citealt{2013ApJ...778..178H,2018ApJ...853L...3H}). Thus, we should keep in mind that the radius $R_e$ here implies the radius of the SMS core and not that of entire SMS.

\subsection{Nuclear Burning}

To determine a realistic SMS core model, we employ the condition that SMSs radiate at Eddington luminosity, which is written as
\beq
L_\mathrm{Edd}\approx 1.248 \times 10^{43}Y_e^{-1}\left({M \over 10^5M_\odot}\right)f_\beta\,\mathrm{erg/s}, \label{Ledd}
\eeq
where $f_\beta$ denotes a geometrical correction which is unity in spherical symmetry and is smaller than unity for rotating SMSs~\citep{1999ApJ...526..937B}. 
Specifically, we introduce the condition that the energy generation rate by nuclear burning is equal to the Eddington luminosity. In this paper, we consider the cold and hot Carbon-Nitrogen-Oxygen (CNO) cycles~\citep{2012ApJ...749...37M}, triple alpha reaction~\citep{1990sse..book.....K}, and reactions between helium and carbon and between helium and oxygen~\citep{1990sse..book.....K} as the nuclear reactions because these are dominant channels. For these nuclear reactions, we employ the specific energy generation rate (in units of erg/s/g) as follows: 
\beqn
\dot \varepsilon_\mathrm{CCNO}&=&4.4\times 10^{25} X_\mathrm{CNO}X_\mathrm{p}\rho 
\left(T_9^{-2/3}e^{-15.231T_9^{-1/3}} \right. \nonumber\\
&&~~~~~~~~~~\left.+8.3\times 10^{-5}T_9^{-3/2}e^{-3.0057/T_9}
\right),\\
\dot \varepsilon_\mathrm{HCNO}&=&4.6 \times 10^{15} X_\mathrm{CNO}, \\
\dot \varepsilon_\mathrm{3\alpha}&=&5.09\times 10^{11}\rho^2 X_\mathrm{He}^3 T_8^{-3} e^{-44.027/T_8}, \\
\dot \varepsilon_\mathrm{C+He}&=&1.3 \times 10^{27}X_\mathrm{He} X_\mathrm{C} \rho T_8^{-2} \left({1+0.134T_8^{2/3} \over 1+0.017 T_8^{2/3}}\right)^2 
\nonumber \\
&&~~~~\times e^{-69.20/T_8^{2/3}},
\\
\dot\varepsilon_\mathrm{O+He}&=&X_\mathrm{He} X_\mathrm{O} \rho 
\left[T_9^{-2/3} e^{-39.76T_9^{-1/3}-(T_9/1.6)^2} \right.
\nonumber \\
&&~~~~~~~~~~~\left.+3.64\times 10^{18}T_9^{-3/2}e^{-10.39/T_9} \right.
\nonumber \\
&&~~~~~~~~~~~\left.+4.39\times 10^{19} T_9^{-3/2}e^{-12.2/T_9} \right.
\nonumber \\
&&~~~~~~~~~~~\left.+2.29\times 19^{16}T_9^{2.966}e^{-11.9/T_9}\right].
\eeqn
Here, $T_k=T/(10^k\,\mathrm{K})$, the units of $\rho$ are $\mathrm{g/cm}^3$, and $X_\mathrm{CNO}$ is the mass fraction of CNO elements. 
We ignore the screening factor and small correction factors in this paper. Note that when the energy generation rate by the CNO cycle is considered, we define
\beq
\dot\varepsilon_\mathrm{CNO}=\mathrm{min}(\dot \varepsilon_\mathrm{CCNO},\dot\varepsilon_\mathrm{HCNO}).
\eeq
The minimum is taken because above $T_8 \approx 3$ only the hot CNO cycle is efficient, even though the energy generation rate $\dot \varepsilon_\mathrm{CCNO}$ would be higher, and vice-versa for low temperatures. Note also that carbon and oxygen burning play an important role only for high temperatures of $T_8 \agt 8$, for which the SMSs have already experienced the pair instability (see Sec.~\ref{secIIID}). 

The energy generation rate is calculated by
\beq
\dot E = 2\pi c \iint \rho u^t B e^{2\zeta-2\nu} \dot \varepsilon \,r^2 dr d(\cos\theta),
\eeq
where $\dot \varepsilon=\dot\varepsilon_\mathrm{CNO}+ \dot\varepsilon_{3\alpha}+\dot\varepsilon_\mathrm{C+He}+\dot\varepsilon_\mathrm{O+He}$. A stellar model is determined from the condition of $\dot E=L_\mathrm{Edd}$. 
Specifically, for a given value of $\Gamma$, we search for a solution with a particular value of the central density $\rho_c$ (and hence central temperature $T_c$ for the given equation of state) which satisfies this equality. The desired mass is obtained by varying $\Gamma$: As Eq.~(\ref{mass20}) indicates, the mass is determined approximately by $\sigma Y_T$, and thus, for a given composition, the desired mass is easily obtained by controlling $\sigma$. For spherically symmetric and slowly rotating SMS cores ($\beta\alt 0.004$), we simply employ $f_\beta=1$ while for SMS cores at mass shedding limit, we employ $f_\beta=2/3$, approximately following the result of \cite{1999ApJ...526..937B}. 

\subsection{Analysis for Stability}

The stability against the general relativistic instability is determined by the following fitting formula derived in our previous paper~\citep{Shibata:2016vxy}:
\beq
F:=\Gamma-{4 \over 3}  
-\left[y-y^2 - \left({10 \over 3}-2 \Gamma - y -\beta\right)\beta\right],\label{final}
\eeq
where $\beta$ and $y$ are defined in Eqs.~(\ref{eq:beta-rotation-parameter}) and (\ref{eq:dimensionless-quantity-y}), respectively. For $F < 0$, the SMS cores are unstable against the radial oscillation of the fundamental mode, leading to stellar collapse. In the previous paper, we showed that this criterion is satisfied for a wide range of $\Gamma$ of $4/3 \leq \Gamma \leq 1.34$, which is valid for the SMS cores considered in this paper. 

We note that for $\beta=0$ and $y \ll 1$, the relation of $F=0$ is written as $1/(6\sigma)=P_\mathrm{c}/(\rho_\mathrm{c}c^2) \approx aT_\mathrm{c}^4/(3\rho_\mathrm{c}c^2)$. Because of Eq.~(\ref{sigmadef}), we have $\sigma \propto T_\mathrm{c}^3/(\rho_\mathrm{c}Y_T)$, and thus, along the marginally stable curve of $F=0$, $\rho_\mathrm{c}\propto T_\mathrm{c}^{7/2} Y_T^{-1/2}$. Using Eq.~(\ref{mass20}), we also obtain a well-known relation for marginally stable spherical SMS cores as $\rho_\mathrm{c} \propto Y_T^3 M^{-7/2}$~\citep{1983bhwd.book.....S,1986ApJ...307..675F}. For the rotating case with $\beta \sim y$, these relations are not satisfied because in this case, $F=0$ approximately gives $1/(6\sigma)=P_\mathrm{c}/(\rho_\mathrm{c}c^2)-2\beta/3$. In Sec.~\ref{secIIID} we show that the correction by $\beta$ is appreciable for $\beta \agt 10^{-3}$ in determining the instability region. 

At high temperatures of $T \agt 5 \times 10^8$\,K, the creation of $e^- e^+$ pairs softens the equation of state. We analyse the equation of state for such high-temperature regions by modifying Eq.~(\ref{pressure}) and 
identify a region in the plane of the density and temperature for which the adiabatic index becomes less than $4/3$. If the central density and temperature for a star are in this region, we consider that such a star is unstable against the pair instability.

\begin{figure*}[t]
\includegraphics[width=0.49\textwidth]{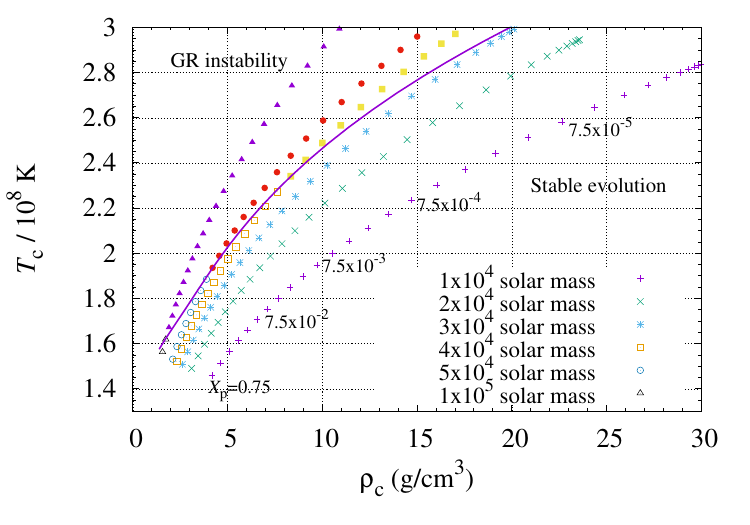}
\includegraphics[width=0.49\textwidth]{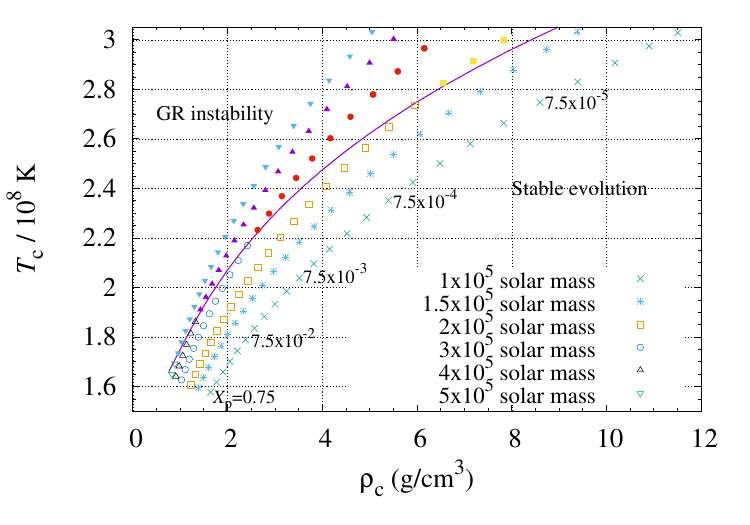}
\vspace{-4mm}
\caption{Equilibrium sequences of SMS cores in the hydrogen burning phase composed of hydrogen, helium, electrons, photons, and a small fraction of CNO elements with fixed values of the gravitational mass in the plane of the central rest-mass density, $\rho_\mathrm{c}$, and temperature, $T_\mathrm{c}$. The left and right panels show the spherical and mass-shedding limit cases, respectively. The mass fraction of CNO elements is assumed to be $5 \times 10^{-9}$ for these plots. 
The sequences are constructed with a varying mass fraction of hydrogen as $X_\mathrm{p}=0.75 \times 10^{-0.2i}$ with $i=0, 1, 2, \cdots$. The open and filled symbols along each sequence that crosses the solid curve denote the stable and unstable SMS cores: The solid curve shows the threshold of the general relativistic instability ($F=0$), above which temperature the SMS cores are unstable. 
}
\label{fig1}
\end{figure*}

\subsection{Results}\label{secIIID}

\subsubsection{Hydrogen burning case}

\begin{figure*}[t]
\includegraphics[width=0.49\textwidth]{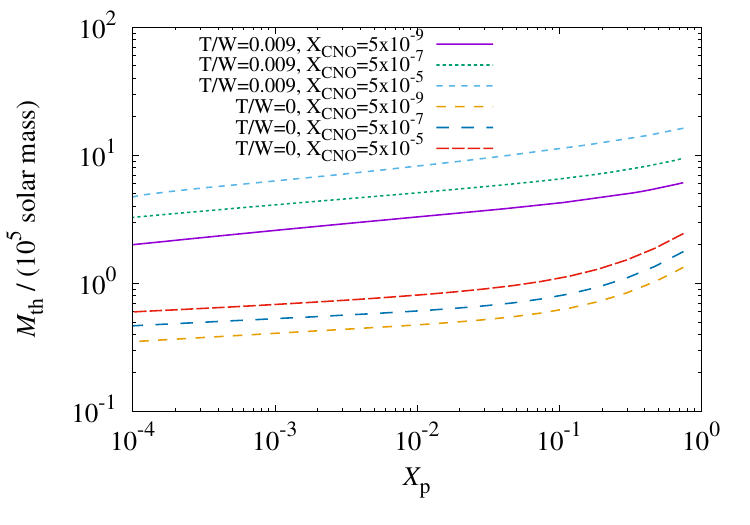}
\includegraphics[width=0.49\textwidth]{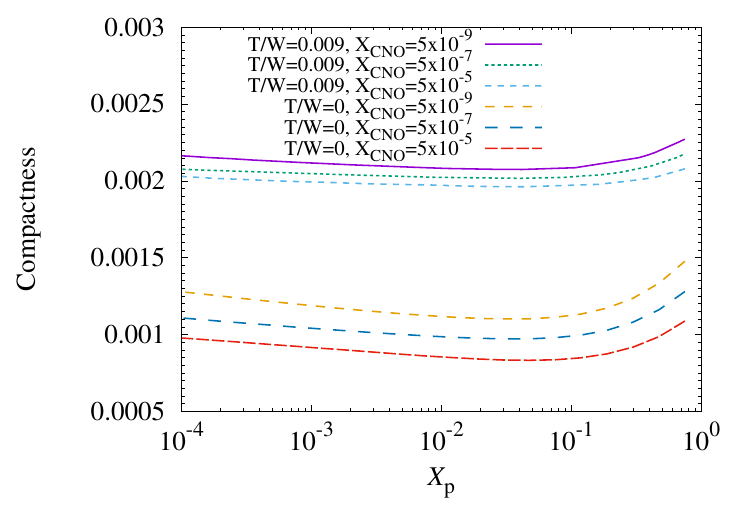}
\vspace{-4mm}
\caption{Left: Threshold mass $M_\mathrm{th}$ of SMS cores in the hydrogen burning phase against the general relativistic instability as a function of the hydrogen mass fraction $X_\mathrm{p}$. We show the cases with $\beta \approx 0.009$, $0$ and for $X_\mathrm{CNO}=5\times 10^{-9}$, $5\times 10^{-7}$, and $5\times 10^{-5}$. 
Right: The same as the left panel but for the compactness of the SMS cores at $M=M_\mathrm{th}$ as a function of $X_\mathrm{p}$.
}
\label{fig1b}
\end{figure*}

\begin{figure*}[t]
\includegraphics[width=0.49\textwidth]{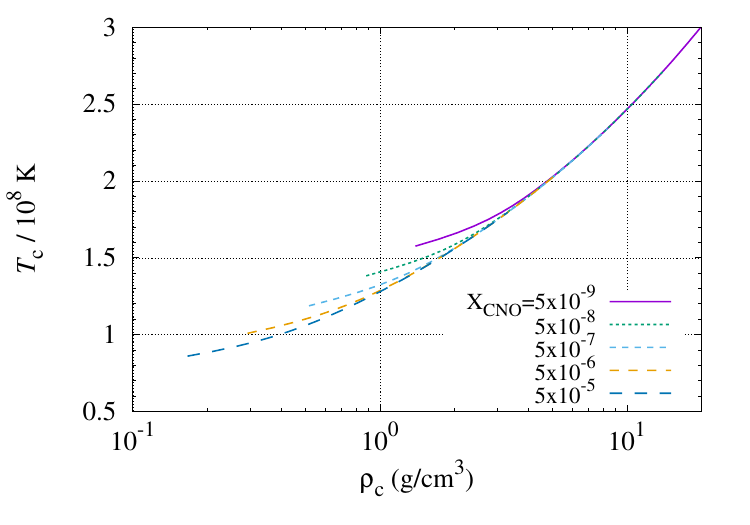}
\includegraphics[width=0.49\textwidth]{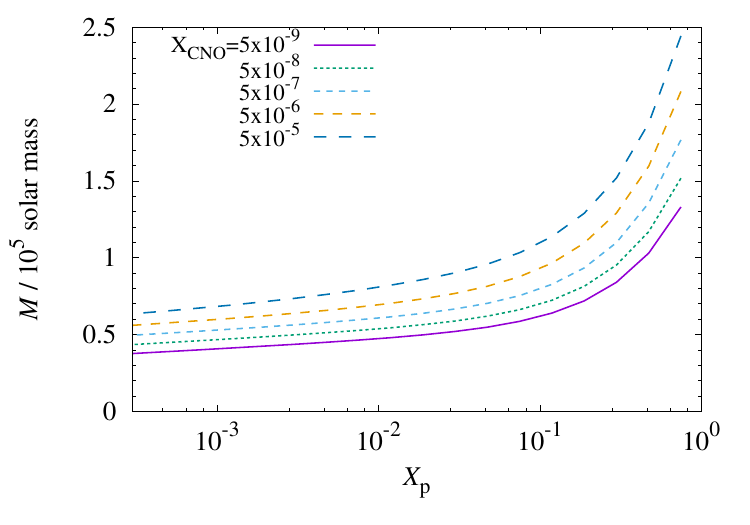}\\
\includegraphics[width=0.49\textwidth]{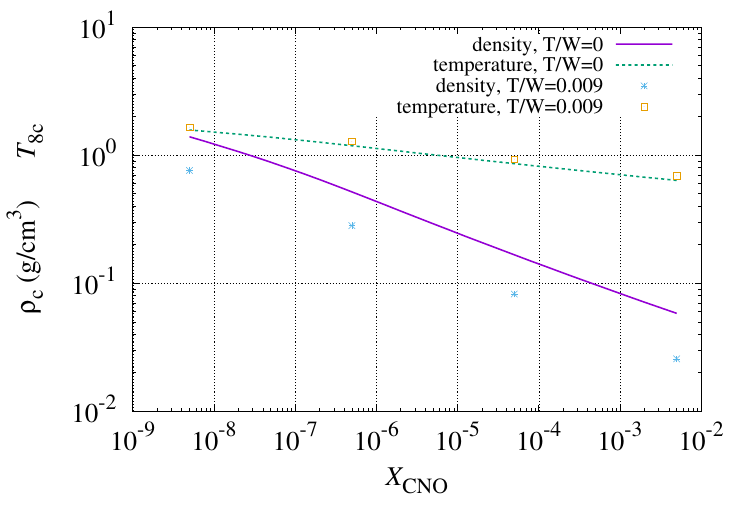}
\includegraphics[width=0.49\textwidth]{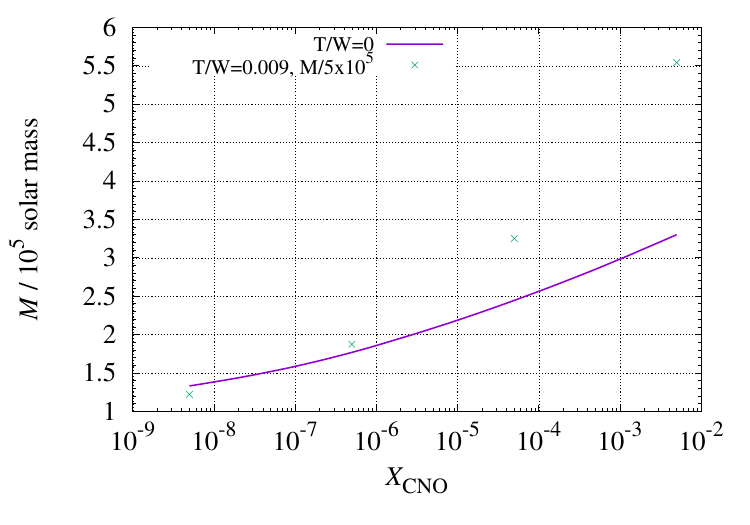}\\
\includegraphics[width=0.49\textwidth]{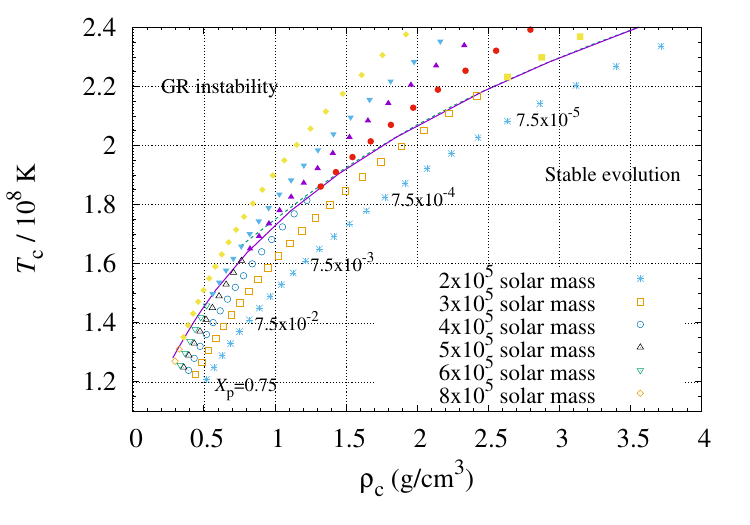}
\includegraphics[width=0.49\textwidth]{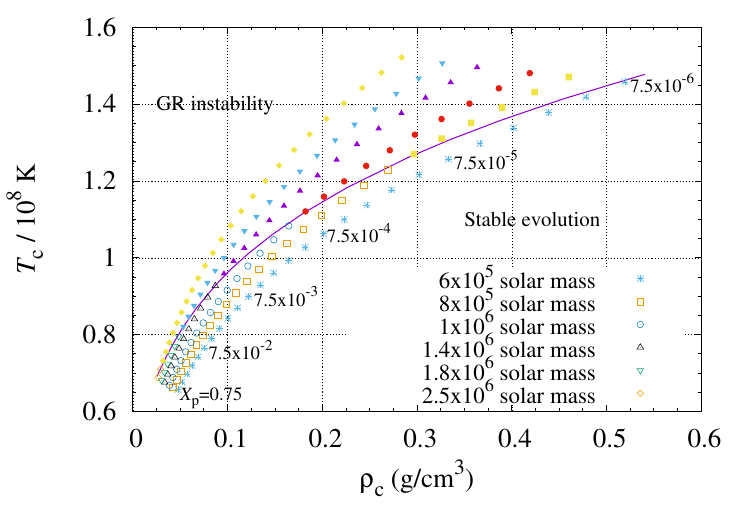}
\vspace{4mm}
\caption{Top left: The relation between $\rho_\mathrm{c}$ and $T_\mathrm{c}$ for spherical SMS cores which are marginally stable to the general relativistic instability. For the lowest central density and temperature, $X_\mathrm{p}=0.75$ and for their higher values, $X_\mathrm{p}$ is smaller. 
Top right: The mass as a function of $X_\mathrm{p}$ for SMS cores marginally stable to the general relativistic instability for different values of $X_\mathrm{CNO}$.
Middle left and right: Central density in units of $\mathrm{g/cm^3}$ and temperature in units of $10^8$\,K as functions of $X_\mathrm{CNO}$ (left) and threshold mass as a function of $X_\mathrm{CNO}$ (right) for spherical SMS cores (curves) and rotating SMS cores with $\beta \approx 0.009$ (symbols) in hydrogen burning phases with $X_\mathrm{p}=0.75$ and different values of $X_\mathrm{CNO}$. For the mass of the rotating SMS cores, we plot $M/(5\times 10^5M_\odot)$ to clarify the high enhancement factor of the threshold mass by the rotation. 
Bottom left and right: The same as the right panel of Fig.~\ref{fig1} but for $X_\mathrm{CNO}=5 \times 10^{-7}$ and $5 \times 10^{-3}$, respectively. The solid and dotted curves show the threshold of the general relativistic instability for given values of $X_\mathrm{CNO}$ and $X_\mathrm{CNO}=5 \times 10^{-9}$, respectively. 
}
\label{fig1c}
\end{figure*}

\begin{figure*}[t]
\includegraphics[width=0.49\textwidth]{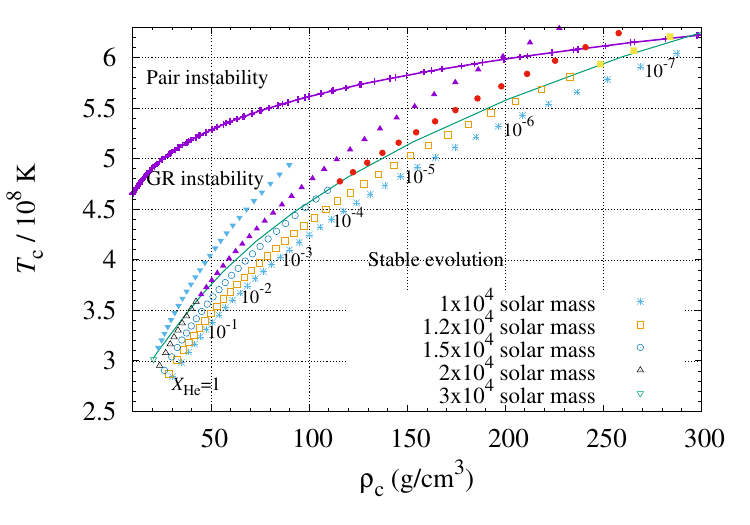}
\includegraphics[width=0.49\textwidth]{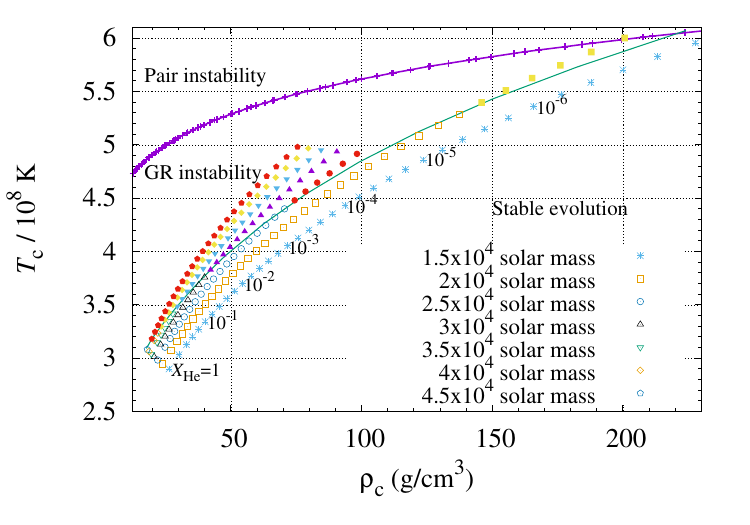}\\
\includegraphics[width=0.49\textwidth]{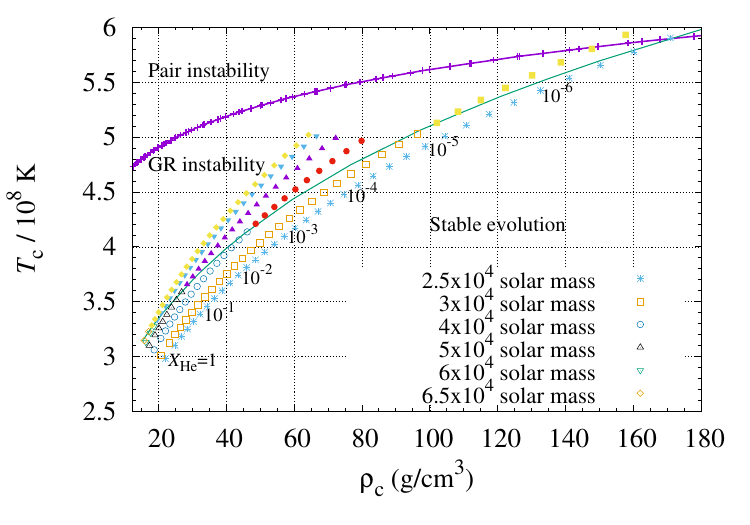}
\includegraphics[width=0.49\textwidth]{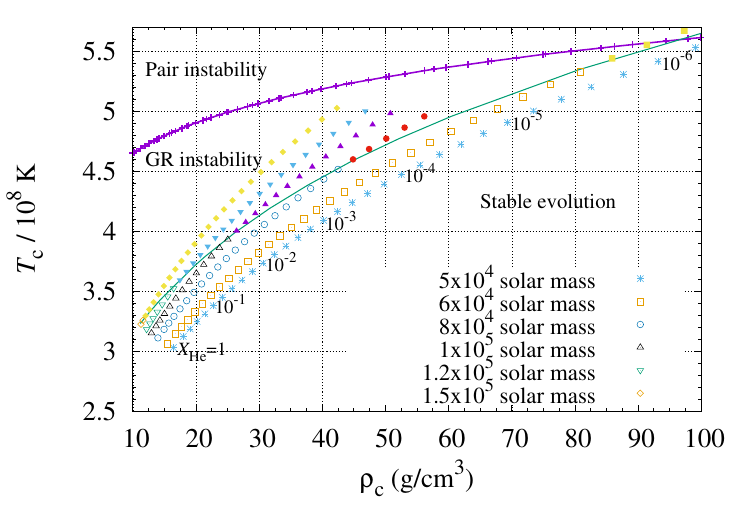}
\vspace{-4mm}
\caption{Equilibrium sequences of SMS cores in the helium burning phase composed of helium, carbon, oxygen, electrons and photons, with fixed values of the gravitational mass, in the plane of the central rest-mass density, $\rho_\mathrm{c}$, and temperature, $T_\mathrm{c}$. The different panels show the spherical case $\beta=0$ (top left), the case $\beta \approx 2.05\times 10^{-3}$ (top right), the case $\beta \approx 4.05\times 10^{-3}$ (bottom left), and the case at the mass shedding limit $\beta \approx 9.0\times 10^{-3}$ (bottom right). The sequences are constructed while varying the mass fraction of helium as $X_\mathrm{He}=10^{-0.2i}$ with $i=0, 1, 2, \cdots$. Specific values of $X_\mathrm{He}$ are shown along each plot at intervals of $X_\mathrm{He}=10^{-i}$. The lower solid curve shows the threshold of the general relativistic instability ($F=0$) above which temperature the SMS cores are unstable:  The open and filled symbols along each sequence that crosses this threshold solid curve denote the stable and unstable SMS cores, respectively. The upper curve with symbols shows the threshold for the pair instability, above which temperature the SMS cores become unstable due to $e^-e^+$ pair production. 
}
\label{fig2}
\end{figure*}

Figure~\ref{fig1} displays equilibrium sequences of SMS cores with fixed values of the mass in the plane of the central density $\rho_\mathrm{c}$ and central temperature $T_\mathrm{c}$ for the hydrogen burning phase. Here the sequence does not strictly imply the evolution sequence because we do not take into account possibly important effects such as mass loss and shell burning. We assume for the mass fraction of CNO elements that $X_\mathrm{C}=X_\mathrm{CNO}=5\times 10^{-9}$ (see~\citealt{Bond:1984sn}). In this plot, we simply assume that $X_\mathrm{He}=1-X_\mathrm{p}-X_\mathrm{CNO}$ (i.e., no further production of CNO elements). Thus, for the models with $T_\mathrm{c} \agt 2.7\times 10^8$\,K, for which the helium burning is activated (cf.~Fig.~\ref{fig2} and also see \citealt{2020MNRAS.496.1224N}), the results are not very reliable. We also note that for low values of $X_\mathrm{p}$, shell burning is likely to set in and the results here would not be very accurate. 

The left and right panels of Fig.~\ref{fig1} show the results for the spherical case and for the mass-shedding limit case ($\beta\approx 0.009$), respectively. The solid curve denotes the threshold of the general relativistic instability ($F=0$). For the higher temperature side above this curve, the SMS core is unstable against gravitational collapse. The open and filled symbols along each sequence that crosses the stability curve denote the stable and unstable stars, respectively. Each sequence is plotted for given mass with $M=10^4$--$10^5M_\odot$ for the spherical case and $1 \times 10^5$--$5\times 10^5M_\odot$ for the mass-shedding case. 

The left panel of Figure~\ref{fig1b} shows the threshold mass for the general relativistic instability as a function of the hydrogen mass fraction. The right panel shows the compactness $\mathcal{C}$ of the SMS cores at the threshold mass as a function of the hydrogen mass fraction. 

Figures~\ref{fig1} and \ref{fig1b} show that for $M \alt 3\times 10^4M_\odot$, the spherical SMS cores before the onset of helium burning at $T_\mathrm{c} \sim 3\times 10^8$\,K are always stable against the general relativistic instability. On the other hand, for higher masses of $3 \times 10^4M_\odot \alt M \alt 1.3\times 10^5M_\odot$, the SMS cores should encounter the instability during their stellar evolution: The threshold mass decreases with the decrease of the hydrogen mass fraction. The reason for this is that the hydrogen burning reduces the value of $Y_T$, and as a result, the value of $\sigma$ is increased ($\Gamma-4/3$ is decreased) for a given value of the stellar mass (see Eq.~(\ref{mass20}) for the dependence of the stellar mass on $Y_T$ and $\sigma$). 


For the rapidly rotating case at the mass-shedding limit, the threshold mass for the general relativistic instability is significantly enhanced (see the left panel of Fig.~\ref{fig1b}). For this case, the SMS cores with $M\alt 2\times 10^5M_\odot$ are stable against the general relativistic instability during the hydrogen burning phase. SMS cores with $2\times 10^5 \alt M \alt 6 \times 10^5M_\odot$ become unstable during the hydrogen burning phase, and the threshold mass decreases with the hydrogen depletion. In other words, the curve of the marginally stable SMS cores in the $\rho_\mathrm{c}$-$T_\mathrm{c}$ plane is shifted to the low-density side for a given value of $T_\mathrm{c}$ (compare both panels of Fig.~\ref{fig1} noting that the horizontal ranges are different). 


The right panel of Fig.~\ref{fig1b} shows the compactness, ${\cal C}$, for given values of $\beta$ and $X_\mathrm{CNO}$. We find that ${\cal C}$ lies in a narrow range irrespective of $X_\mathrm{p}$. For the spherical case with $X_\mathrm{CNO}=5 \times 10^{-9}$, it is between $1.1\times 10^{-3}$ and $1.5\times 10^{-3}$. For the mass shedding limit ($\beta\approx 0.009$) with $X_\mathrm{CNO}=5 \times 10^{-9}$, it is between $2.0 \times 10^{-3}$ and $2.3\times 10^{-3}$. By contrast, its dependence on the rotational degree $\beta$ is larger. Changing $\beta$ significantly affects the compactness and for the rotating case, the compactness is larger. The reason for this is that in the presence of rotation, the self-gravity of the star has to be stronger than that in the non-rotating case to reach the marginally stable state against the general relativistic instability. 

It is also noted that the axial ratio of the polar axis to the equatorial axis is $\approx 2/3$ for the mass shedding limit. Thus, the morphology is quite non-spherical, and indeed, the photon flux evaluated from the spatial derivative of $T^4$ is appreciably smaller for the equatorial direction than for the polar direction (see \citealt{1999ApJ...526..937B}). This implies that the radiation pressure effect on the equatorial plane is much less important than along the polar axis (see, e.g.,  \citealt{2023ApJ...950..184K} for a related study). All these facts indicate that it is important to accurately take into account the non-spherical shape when one studies the evolution of rapidly rotating SMSs. 

For marginally stable SMS cores in rigid rotation, $\beta$ and $q$ are in the ranges of $0 \leq \beta \alt 0.009$ and $0 \leq q \alt 0.85$. Therefore, a black hole with a not-extremely high spin is likely to be an outcome after the collapse as we found in our previous paper~\citep{Uchida2017oct}, if the stellar explosion does not occur during the collapse~\citep{2020MNRAS.496.1224N}. 

The relation between $\rho_\mathrm{c}$ and $T_\mathrm{c}$ for SMS cores marginally stable to the general relativistic instability depends on the CNO mass fraction. The top-left panel of Fig.~\ref{fig1c} shows the curves in the $\rho_\mathrm{c}$-$T_\mathrm{c}$ plane for marginally stable spherical SMS cores with $X_\mathrm{CNO}=5 \times 10^{-k}$ with $k\in\{5,6,7,8,9\}$. For the larger values of $X_\mathrm{CNO}$, the central density and temperature are smaller for a given value of $X_\mathrm{p}$. The middle left panel of Fig.~\ref{fig1c} also shows that the decrease in the central density is more remarkable than in the central temperature. It is worth noting that the central temperature depends only weakly on the degree of rotation for given values of $X_\mathrm{CNO}$, while the central density becomes lower in the presence of rotation. 

The threshold mass for a given value of $X_\mathrm{p}$ is larger for larger values of $X_\mathrm{CNO}$ (see the left panel of Fig.~\ref{fig1b} and the top-right and middle-right panels of Fig.~\ref{fig1c}). In the presence of high mass fractions of CNO elements, the threshold mass for the onset of the general relativistic instability is appreciably enhanced. This is because the fraction of the radiation pressure is enhanced; the value of $\sigma$ is increased (see Eq.~(\ref{sigmadef})). As a result, the value of $\Gamma$ approaches $4/3$. This enhancement is further increased in the presence of rotation because $\Gamma$ for the marginally stable SMS core with rotation is closer to $4/3$ than for the spherical SMS cores. For example, for $X_\mathrm{CNO}=5 \times 10^{-3}$, the threshold mass for $\beta \approx 0.009$ is about 9 times higher than that for the spherical case. 

\begin{figure*}[t]
\includegraphics[width=0.49\textwidth]{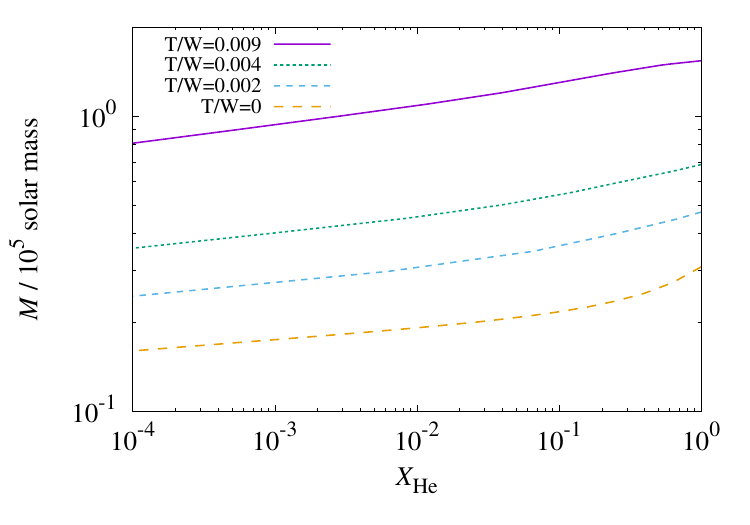}
\includegraphics[width=0.49\textwidth]{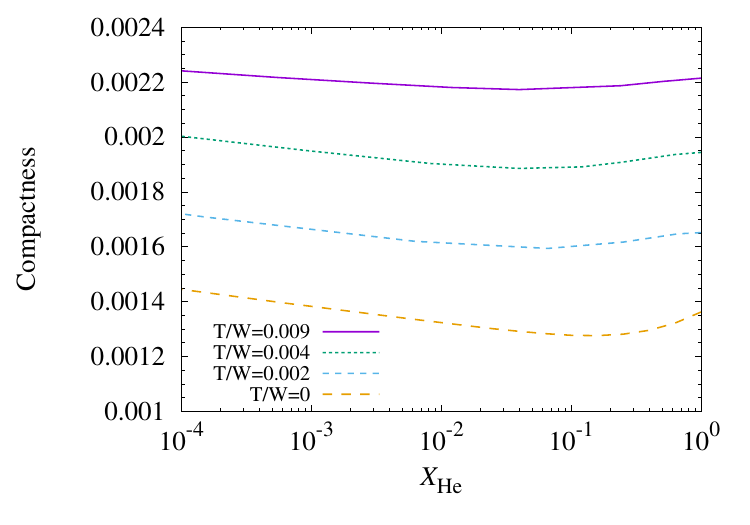}
\vspace{-4mm}
\caption{Left: Threshold mass $M_\mathrm{th}$ for SMS cores in the helium burning phase against the general relativistic instability as a function of the helium mass fraction $X_\mathrm{He}$. 
Right: The same as the left panel but for the compactness at $M=M_\mathrm{th}$ as a function of $X_\mathrm{He}$.
}
\label{fig2b}
\end{figure*}

The bottom two panels of Fig.~\ref{fig1c} show the same plot as the right panel of Fig.~\ref{fig1} but for $X_\mathrm{CNO}=5\times 10^{-7}$ (left) and $5\times 10^{-3}$ (right), where all the SMS cores have $\beta \approx 0.009$. It is clearly seen that the stability threshold curve shifts to the lower density and lower temperature with the increase of $X_\mathrm{CNO}$ (note again that the decrease in the density is more remarkable). This is also the case for the spherical case. Thus, for larger values of $X_\mathrm{CNO}$, the unstable SMS cores have a longer dynamical timescale ($\propto \rho_\mathrm{c}^{-1/2}$) and the collapse starts at lower temperature. This point is important for our discussion in Sec.~\ref{sec4} about possible explosion scenarios after the onset of collapse due to the general relativistic instability.

\subsubsection{Helium burning case}

Figure~\ref{fig2} shows equilibrium sequences of SMS cores in the helium burning phase composed of helium, carbon, oxygen, electrons, and photons. Each sequence has a fixed value of the gravitational mass and $\beta$ is chosen as $\in\{0, ~2.05, ~4.05, ~9.0\} \times 10^{-3}$. The sequences are constructed while varying the mass fraction of helium as $X_\mathrm{He}=10^{-0.2i}$ with $i=0, 1, 2, \cdots$. For simplicity, we assume that the mass fraction of carbon and oxygen is identical for each sequence. Computations were also performed while changing the ratio of this fraction as 2:1, 1:3, and 1:5. Since the value of $Y_T$ is modified, the results change slightly. However, we do not observe a significant difference among these cases. 
The solid curves show the stability threshold $F=0$, above which temperature the SMS cores are unstable to the general relativistic instability. The open and filled symbols along each sequence that crosses the curve of $F=0$ again denote the stable and unstable SMS cores, respectively. The upper curve with symbols shows the threshold for the pair instability, above which temperature the SMSs become unstable due to the $e^-e^+$ pair production. 

Figure~\ref{fig2b} also plots the threshold mass for the general relativistic instability as a function of the helium mass fraction and the compactness of the SMS cores at the threshold mass as a function of the helium mass fraction. 
As in the hydrogen burning phase, the threshold mass for the general relativistic instability decreases as a result of helium depletion. For example, for spherical SMS cores with $1\times 10^4M_\odot \alt M \alt 3\times 10^4M_\odot$, the core is likely to become unstable against the general relativistic instability during the stellar evolution. Only for the relatively low-mass SMS cores with $M\alt 10^4M_\odot$, the collapse to a black hole would be triggered by the pair instability (cf. \citealt{2024arXiv240706994N} for a related topic).  

Again, the threshold mass increases significantly in the presence of rotation. For the mass-shedding limit case, it is about 5 times higher than for the spherical case, irrespective of $X_\mathrm{He}$. 
For the most rapidly rotating case, the SMS cores with $M \agt 6 \times 10^4M_\odot$ become unstable to the general relativistic instability before the onset of the pair instability. Thus, for the rapidly rotating SMS cores, the collapse to a black hole is likely to be triggered by the pair instability even when the mass is relatively high as $M \alt 6 \times 10^4M_\odot$. 



The compactness of the marginally stable helium burning SMS cores is in a range similar to that of the hydrogen burning ones with $X_\mathrm{CNO}=5 \times 10^{-9}$; for the spherical case it is between $1.3\times 10^{-3}$ and $1.6\times 10^{-3}$. For larger values of $\beta$, the compactness increases. The resulting values are similar to those for the hydrogen burning SMS cores.

\begin{figure}[t]
\includegraphics[width=0.49\textwidth]{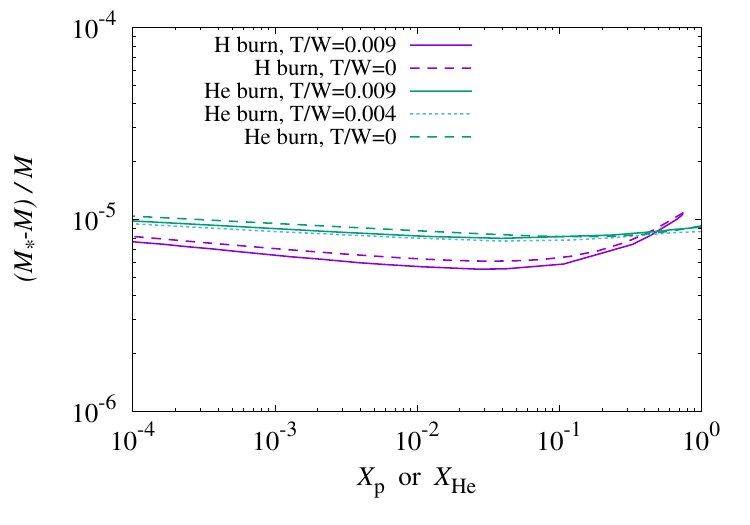}
\vspace{-8mm}
\caption{Normalized binding energy, $(M_*-M)/M$, of SMS cores at the onset of the general relativistic instability. The horizontal axis denotes $X_\mathrm{p}$ in the hydrogen-burning phase with $X_\mathrm{CNO}=5\times 10^{-9}$ or $X_\mathrm{He}$ in the helium burning phase, respectively. 
}
\label{fig3}
\end{figure}

\section{The possibility of an explosion after the onset of collapse}\label{sec4}

Irrespective of the stellar composition and rotation, we find that rigidly rotating SMS cores fulfill the relation for the binding energy, $(M_*-M)/M =\mathcal{O}(10^{-5})$, at the threshold against the general relativistic instability (see Fig.~\ref{fig3}). The value of $(M_*-M)/M$ becomes even smaller, down to $\sim 0.5\times10^{-5}$, for the SMS cores in the hydrogen burning phase with large values of $X_\mathrm{CNO}$. This is because $\Gamma$ approaches $4/3$ for this case. The finding here implies that the absolute value of the binding energy of SMS cores defined by $|M-M_*|$ is two order of magnitude smaller than $GM^2/R_\mathrm{e}=M c^2 {\cal C}$. By contrast, $W/(Mc^2)$ and $U/(Mc^2)$ are much larger than this as $2\times 10^{-3}$--$5\times 10^{-3}$ with $W \agt U \gg T_\mathrm{rot}$ and $(W-U)/(Mc^2)=\mathcal{O}(10^{-5})$. For more rapidly rotating cases, the values of $W/Mc^2$ and $U/Mc^2$ are higher. The small binding energy and the relation of $W \approx U$ reflect the nature of SMS cores for which $\Gamma$ is close to 4/3~\citep{1971reas.book.....Z}. The analysis here shows that, to get an explosion of SMSs after the start of the collapse due to the general relativistic instability, an energy injection at least of order $10^{-5}Mc^2$ is necessary during the collapse. 

Based on these results, we consider the possibility that a stellar explosion occurs after the onset of collapse due to the general relativistic instability. 
As shown in \cite{Uchida2017oct}, an energy injection of $\agt 10^{-5}Mc^2$ is possible from a massive disk that experiences a strong shock, and the explosion of the outer layer of SMSs could be achieved. Another possibility is energy injection from explosive nuclear burning. Here we consider this second possibility. 

Just before the onset of the collapse, the luminosity, and thus, the energy generation rate by the nuclear burning is given by Eq.~(\ref{Ledd}). The collapsing time until the formation of a black hole is approximately estimated by the free-fall timescale as
\beqn
t_\mathrm{ff}&=&(G\bar\rho)^{-1/2} \nonumber \\
&=&1.7\times 10^4\,{\rm s} \left({{\cal C} \over 1.5\times 10^{-3}}\right)^{-3/2}
\left({M \over 10^5M_\odot}\right),
\eeqn
where $\bar\rho$ is the average density and we assumed spherical symmetry in this analysis for simplicity. Then, in the case where the nuclear energy generation rate is equal to the Eddington luminosity, we have 
\beqn
\Delta E&=&t_\mathrm{ff} L_\mathrm{Edd}=2.2 \times 10^{47} Y_e^{-1} f_\beta \,{\rm erg}
\nonumber \\
&&~~~~\times \left({{\cal C} \over 1.5\times 10^{-3}}\right)^{-3/2}
\left({M \over 10^5M_\odot}\right)^2,
\eeqn
and thus,
\beqn
{\Delta E \over Mc^2}&=&1.2 \times 10^{-12} Y_e^{-1} f_\beta \nonumber \\
&&~~\times \left({{\cal C} \over 1.5\times 10^{-3}}\right)^{-3/2}
\left({M \over 10^5M_\odot}\right). \label{eq35}
\eeqn
This implies that a significant enhancement of the nuclear burning efficiency by, at least, a factor $\sim 10^7$ is required for the SMS core to explode during the collapse. We here note that during the late stage of the collapse, neutrino cooling is significantly enhanced~\citep{2012ApJ...749...37M,Uchida2017oct}. Therefore, the required enhancement is even higher just prior to the black hole formation.  

The requirement may be stated in a different way: Since the binding energy of the system is always $\sim 10^{-5}Mc^2$, the required input of specific energy is $10^{-5}c^2 \approx 9\times 10^{15}$\,erg/g. During the collapse, such energy input (including the loss by the neutrino emission) is necessary for the star to explode. A more detailed analysis will be presented in a future paper. 

For SMS cores in the helium burning phase, such an enhancement is unlikely to be achieved until the onset of the pair instability,  because the nuclear burning rate is enhanced only by $\sim 10^4$ times when the temperature increases from $\sim 3 \times 10^8$\,K to $\alt 6\times 10^8$\,K. At higher temperature, the pair instability sets in, and the density, which is approximately proportional to $T^3$ in the isentropic condition, increases at most by a factor of 10. Thus, for these stars, the explosion could be achieved only after the central region enters into the pair-unstable region. Although for such high temperature for which the pair creation of $e^-e+$ appreciably depletes the thermal energy, the O$+$He reaction (not the triple He reaction nor C$+$He reaction) is significantly enhanced~\citep{1990sse..book.....K}, and the explosion may be triggered as demonstrated in \cite{2014ApJ...792...44C}. It is interesting to explore this possibility for the explosion in the presence of an appreciable stellar rotation.


For SMS cores in the hydrogen burning phase, the nuclear energy generation rate by the CNO cycle is enhanced only until the temperature reaches $\sim 3\times 10^8$\,K. At this temperature, the hot CNO cycle, for which the nuclear burning rate is independent of the temperature, becomes dominant. For $X_\mathrm{CNO}=5 \times 10^{-9}$, the central temperature at the onset of the general relativistic instability is $\sim 1.6\times 10^8$\,K. The enhancement of the nuclear energy generation rate to $T_c\sim 3\times 10^8$\,K is therefore too low to lead to an explosion. In the presence of sufficient CNO elements before the collapse, the central temperature and density can be quite low as shown in Sec.~\ref{secIIID}. For such a case, the energy generation rate by the CNO cycle is significantly enhanced by the increase of the central temperature before $T_c\sim 3\times 10^8$\,K is reached, leading to the stellar evolution~\citep{1973ApJ...183..941F,1986ApJ...307..675F, Nagele:2024aev}. For the rotating SMS cores, the density at the onset of the general relativistic instability is even lower, i.e., the free-fall timescale is longer, and hence, they are more subject to the explosion. However, the required mass fraction of the CNO elements is extremely high. Thus, for the collapse of SMS cores in the hydrogen burning phase with a plausible metallicity, the final fate is likely to be a black hole. 

Before closing this section, we note that the condition of $\Delta E \agt 10^{-5} Mc^2$ for the explosion is a necessary condition, not the sufficient condition. This can be found by analyzing the virial theorem. For simplicity we here consider the following virial theorem in Newtonian gravity (e.g., chapter 7 of \citealt{1983bhwd.book.....S}) to semi-qualitatively understand the additional condition for the explosion:
\beq
{1 \over 2} \int d^3x \rho \,{d^2 r^2 \over dt^2} =2T_\mathrm{kin} - W + 3\Pi,
\eeq
where $T_\mathrm{kin}$ is the kinetic energy and $\Pi$ is the volume integral of the pressure. Note that the left hand side is different from the usual form because the rest mass does not conserve in this problem. For a SMS core with an equation of state close to a $\Gamma=4/3$ polytrope, $\Pi \approx U/3$. Using the relation of the binding energy in Newtonian gravity, $E=T_\mathrm{kin}-W+U$, we then have
\beq
{1 \over 2}\int d^3x\, \rho \,{d^2 r^2 \over dt^2} \approx T_\mathrm{kin} + E,
\label{eq37}
\eeq
or
\beq
\int d^3x\, \rho r \,{d^2 r \over dt^2} \approx T_\mathrm{kin}-2 T_\mathrm{kin,r} + E,\label{eq38}
\eeq
where $T_\mathrm{kin,r}$ denotes the kinetic energy associated with the radial motion. 
We note that $E$ is negative with $|E|/(Mc^2)=\mathcal{O}(10^{-5})$ initially but in the presence of an efficient nuclear burning it can be positive as we already mentioned and for the infall motion dominant case, $T_\mathrm{kin}-2T_\mathrm{kin,r} \approx - T_\mathrm{kin,r}$

%
During the collapse $d r^2/dt$ is negative. Thus, to get the explosion with $dr^2/dt >0$ (or $dr/dt >0$), $d^2r^2/dt^2$ (or $d^2r/dt^2 > 0$) has to become positive before the formation of a black hole. Equation~(\ref{eq38}) indicates that this can be realized when $E$ becomes positive and $E > T_\mathrm{kin,r}$. 
Since $T_\mathrm{kin}$ is always positive, Eq.~(\ref{eq37}) appears to show that $T_\mathrm{kin}$ may be also important for the bounce. This can be the case when $T_\mathrm{rot}$ is dominant in $T_\mathrm{kin,r}$. However it should not significantly decelerate the collapse because $T_\mathrm{kin}$ stems primarily from the collapsing motion. Thus, to get the explosion, $E$ has to be positive and at least larger than $T_\mathrm{kin,r}$. 
Furthermore the timescale defined by $|(dr/dt)/(d^2r/dt^2)|$ has to become shorter than the dynamical timescale to invert the collapse to the explosion. As the discussion here shows, additional conditions, besides $\Delta E > 10^{-5}Mc^2$, are necessary to get the explosion. 

\section{Summary}\label{sec5}

We studied the dependence of the final fate of SMS cores on their mass and angular momentum. It is found that spherical SMS cores in the hydrogen burning phase encounter the general relativistic instability during the stellar evolution if the mass is larger than $\sim 3 \times 10^4M_\odot$. For rapidly rotating SMS cores, this threshold mass is enhanced by up to a factor of $\sim 5$. Spherical SMS cores in the helium burning phase encounter the general relativistic instability prior to the onset of the $e^-e^+$ pair instability if the mass is larger than $\sim 1\times 10^4M_\odot$, and this threshold mass is also enhanced by up to a factor of $\sim 5$ in the presence of rotation. Here the factor depends on the mass fraction of helium $X_\mathrm{He}$. During the stellar evolution, the threshold mass decreases because the value of $Y_T$ monotonically decreases with the nuclear burning.

SMS cores that are stable to the general relativistic instability become eventually unstable against the pair instability. This instability is likely to be encountered after the helium burning phase. It is worthy to note that for rapidly rotating cases, SMS cores with high mass up to $M \sim 6 \times 10^4M_\odot$ are subject to the pair instability.  

It is found that after the onset of the general relativistic instability, SMS cores in the hydrogen burning phase are likely to collapse to a black hole. This is because the enhancement of the nuclear burning efficiency is unlikely to be high enough to halt or reverse the collapse. On the other hand, as previous works demonstrated, SMS cores in the helium burning phase may explode with no black hole formation~\citep{2014ApJ...792...44C, 2020MNRAS.496.1224N, Nagele2022dec} if the O$+$He reaction is significantly enhanced. 

For the collapse of rotating SMSs to a black hole, a disk/torus is likely to be formed around the black hole~\citep{2002ApJ...572L..39S,2007PhRvD..76h4017L,2012ApJ...749...37M,Uchida2017oct}. 
As demonstrated in \cite{2007PhRvD..76h4017L,Uchida2017oct}, at the surface of the disk/torus, shocks are generated and propagate outwards. \cite{Uchida2017oct} shows that the kinetic energy of the shocks is quite large as $10^{-5}$--$10^{-4}Mc^2$, and hence, the collapsed SMSs could shine as ultra-long supernovae. This scenario is possible irrespective of the mass of the SMSs as long as they have a certain amount of angular momentum and they are subject to the general relativistic instability during any stage of the stellar evolution. 

On the other hand, \cite{2014ApJ...792...44C,2020MNRAS.496.1224N} show that SMSs with a particular mass can explode after the onset of the general relativistic instability. The explosion energy in this case is also quite high $\sim 10^{56}$\,erg, and hence, it can also produce an ultra-long supernova. Previous studies investigate the explosion only for spherical SMSs. Exploring the fate of unstable rotating SMSs is an issue left for a future work. 

Deriving light curve model and spectrum for the ultra-long supernovae is an important issue. In a separate paper, we will present the results based on our models of the ultra-long supernovae (Jockel et al. in preparation).

\acknowledgements

This work was in part supported by Grant-in-Aid for Scientific Research (grant Nos.~20H00158 and 23H04900) of Japanese MEXT/JSPS.


\bibliography{reference}

\end{document}